\documentclass[twocolumn,superscriptaddress,groupedaddress,nofootinbib]{revtex4}  
\usepackage{graphicx}  
\usepackage{dcolumn}   
\usepackage{bm}        
\usepackage{amssymb}   
\usepackage{amsmath}
\usepackage{color}
\begin{document}

\title{Constraining Born-Infeld-like Nonlinear Electrodynamics Using Hydrogen's
Ionization Energy}

\author{P. Niau Akmansoy}
\email{pniau7@gmail.com}
\affiliation{Departamento de F\'{\i}sica Te\'{o}rica e Experimental, \\ Universidade Federal do Rio Grande do Norte. Campus Universit\'{a}rio, s/n - Lagoa Nova, CEP 59072-970, Natal, Brazil }

\author{L. G. Medeiros}
\email{leogmedeiros@ect.ufrn.br}
\affiliation{Instituto de F\'{\i}sica Te\'{o}rica, \\ Universidade Estadual Paulista. Rua Bento Teobaldo Ferraz 271 Bloco II, P.O. Box 70532-2, CEP 01156-970, S\~{a}o Paulo, SP, Brazil. }
\affiliation{Escola de Ci\^{e}ncia e Tecnologia, \\ Universidade Federal do Rio Grande do Norte. Campus Universit\'{a}rio, s/n - Lagoa Nova, CEP 59078-970, Natal, RN, Brazil. }


\bigskip
\date{\today}

\begin{abstract}
In this work, the hydrogen's ionization energy was used to constrain the free
parameter $b$ of three Born-Infeld-like electrodynamics namely Born-Infeld
itself, Logarithmic electrodynamics and Exponential electrodynamics. An
analytical methodology capable of calculating the hydrogen ground state energy
level correction for a generic nonlinear electrodynamics was developed. Using
the experimental uncertainty in the ground state energy of the hydrogen atom,
the bound $b>5.37\times10^{20}K\frac{V}{m}$,
where $K=2$, $4\sqrt{2}/3$ and $\sqrt{\pi}$ for the Born-Infeld,
Logarithmic and Exponential electrodynamics respectively, was established. In
the particular case of Born-Infeld electrodynamics, the constraint found for
$b$ was compared with other constraints present in the literature.
\end{abstract}

\maketitle

\section{Introduction}

Nonlinear electrodynamics (NLED) are extensions of Maxwell's electromagnetism
which arise when self-interaction in field equations is allowed. From the
axiomatic point of view, they can be built from a Lagrangian of a vector field
that respects three conditions: invariance under the Lorentz group, invariance
under the $U(1)$ gauge group and the Lagrangian depending only on combinations
of the field and its first derivative, i.e. $\mathcal{L=L}\left(  A_{\mu
},\partial_{\nu}A_{\mu}\right)  $.

The first two NLED proposals emerged in the 1930s in two very different
contexts. In 1934, M. Born and L. Infeld proposed the Born-Infeld
electrodynamics (BI) in order to deal with the divergence of the self-energy
of a point charge \cite{BornInfeld,BornInfeld1}. The BI electrodynamics was
originally conceived as a fundamental theory for electromagnetism, but later
it was found that it was not renormalizable and therefore should be considered
as an effective theory\footnote{This kind of approach was explicitly performed
when E. S. Fradkin and A. A. Tseytlin showed that BI electrodynamics appears
as an effective theory of low energies in open string theories \cite{FraTse}%
.}. In 1936, W. Heisenberg and H. Euler showed that, for energies below the
electron mass, the self-coupling of the electromagnetic field induced by
virtual pairs of electron-positrons can be treated as an effective field
theory \cite{HeisenbergEuler}. This theory is known as Euler-Heisenberg
electrodynamics and it provided the first description of the vacuum
polarization effect present in the QED \cite{Dunne2004}.

Due to different motivations, other nonlinear electrodynamics were proposed - e.g. Logarithmic
and Exponential electrodynamics \cite{Altshuler,Soleng,Hendi,Hendi2} - and the NLED
became a class of electromagnetic theories \cite{Plebanski}. This class of
theories has applications in several branches of physics being particularly
interesting in systems where the NLED are minimally coupled with
gravitation as in the cases of charged black holes \cite{BH1,BH4,BH5,BH6,BH7,BH8,BH9}
and cosmology \cite{Cosm1,Cosm2,Cosm4,Cosm5,Cosm6}.

Nonlinear electrodynamics have some different features with respect to
Maxwell's electrodynamics. Among these features, the most interesting is its
non-trivial structure for radiation propagation. Due to nonlinearity of the
field equations, the electromagnetic field self-interacts generating
deformities in the light cone \cite{CaLeoPJ}. Thus, in the NLED context, the
introduction of a background field affects the propagation velocity of the
electromagnetic waves and generates the birefringence phenomenon. This
phenomenon is present in all physically acceptable NLED with the exception of
BI electrodynamics \cite{Boilat}.

Excluding the Euler-Heisenberg electrodynamics and its variations
\cite{Dunne2004}, all other NLED have at least one free parameter which must
be experimentally constrained \cite{Fouche}. These constraints can be directly obtained from
measurements of atomic transitions \cite{Soff,CarleyKiess} and photon-photon
scattering \cite{Atlas 2017,Ellis} associated with self-interaction of NLED.
Another possibility occurs in the astrophysical context where bounds to the
NLED are imposed through photon splitting process present in magnetars spectra
\cite{Davila2014}. Moreover, for NLED where the birefringence effect is not
negligible, bounds can be established through measurements of vacuum magnetic
birefringence generated by the passage of a polarized laser beam through a
magnetic dipole field (PVLAS collaboration - see \cite{PVLAS} and references therein).

The purpose of this paper is to build a procedure capable of constraining
nonlinear electrodynamics and to apply this procedure to three
Born-Infeld-like electrodynamics. In section \ref{sec - NLED}, an introduction
to the NLED is presented with emphasis on three specific nonlinear
electrodynamics: Born-Infeld NLED, Logarithmic NLED and Exponential NLED. The
procedure based on the hydrogen's ionization energy which constrains NLED is
developed in section \ref{sec - HIE}. In this section, bounds on the free
parameters of each NLED  are  established and the results obtained are
compared with those present in the literature. The final remarks are discussed
in section \ref{sec - Final}.

\section{Nonlinear Electrodynamics\label{sec - NLED}}

The nonlinear electrodynamics in vacuum are described by the Lagrangian
\begin{equation}
\mathcal{L=L}\left(  F,G\right)  , \label{Lagragian}%
\end{equation}
where%
\begin{eqnarray*}
F  &  = & -\frac{1}{4}F^{\mu\nu}F_{\mu\nu}=\frac{1}{2}\left(  E^{2}-B^{2}\right)
,\\
G  &  = & -\frac{1}{4}F^{\mu\nu}\tilde{F}_{\mu\nu}=\vec{E}\cdot\vec{B},
\end{eqnarray*}
are the contractions of the electromagnetic field strength tensor $F^{\mu\nu}$
with its dual $\tilde{F}^{\mu\nu}=\frac{1}{2}\varepsilon^{\mu\nu\alpha\beta
}F_{\alpha\beta}$. The variation of (\ref{Lagragian}) with
respect of $A_{\mu}$ and Bianchi identity result in the nonlinear field
equations%
\begin{eqnarray}
\partial_{\mu}h^{\mu\nu}=\partial_{\mu}\left(  \mathcal{L}_{F}F^{\mu\nu
}+\mathcal{L}_{G}\tilde{F}^{\mu\nu}\right)   &  = & 0,\label{FE1}\\
\partial_{\gamma}F_{\mu\nu}+\partial_{\nu}F_{\gamma\mu}+\partial_{\mu}%
F_{\nu\gamma}  &  = & 0, \label{FE2}%
\end{eqnarray}
where $\mathcal{L}_{F}$ e $\mathcal{L}_{G}$ are the Lagrangian partial
derivatives with respect to the invariants. This set of equations completely
describe the system. The electric system displacement vector, from which
nonlinear effects can be interpreted as a polarization of the medium, are
given by $D^{i}\equiv h^{0i}$ or, in terms of the Lagrangian derivatives, by%

\begin{equation}
\vec{D}=\mathcal{L}_{F}\vec{E}+\mathcal{L}_{G}\vec{B}. \label{D vector}%
\end{equation}

Usually, the system of equations (\ref{FE1}) and (\ref{FE2}) is very difficult
to be analytically solved. An exception is the electrostatic case where
$\vec{E}$ only depends on one variable. In this situation, (\ref{FE2}) is
automatically satisfied and (\ref{FE1}) reduces to%
\begin{equation}
\vec{\nabla}\cdot\vec{D}=0\text{.} \label{FE1 electrostatic}%
\end{equation}
Since the solution of (\ref{FE1 electrostatic}) is identical to the Maxwell
case, the problem becomes an algebraic problem associated with the inversion
of the equation%
\begin{equation}
\vec{D}=\mathcal{L}_{F}\left(  E^{2}\right)  \vec{E}. \label{D and E}%
\end{equation}

\subsection{Born-Infeld-like electrodynamics}

An important sub-class of the NLED arises when (\ref{Lagragian}) is an
analytical function of the $F$ and $G$. In this case, the Lagrangian can be
written as a series of the invariants
\begin{equation}
\mathcal{L=}%
{\displaystyle\sum_{m,n}}
a_{m,n}F^{m}G^{n}=F+a_{2,0}F^{2}+a_{0,2}G^{2}+a_{1,1}FG+..., \label{Series}%
\end{equation}
where the linear coefficient in $G$ can be neglected because of Bianchi
identity. The main NLED (Born-Infeld, Euler-Heisenberg, etc) have this
structure. For instance, the first coefficients for Born-Infeld
electrodynamics \cite{BornInfeld} are%
\begin{equation}
a_{2,0}=a_{0,2}=\frac{1}{2b^{2}}\text{ \ and \ }a_{1,1}=0\text{.}
\label{BI coef}%
\end{equation}
Any NLED which can be expanded as (\ref{Series}) with the first coefficients
given by (\ref{BI coef}) is said a Born-Infeld-like electrodynamics
\cite{Helayel1,Helayel2}. Any two Born-Infeld-like NLED are fundamentally
different, but in the weak field limit, when the nonlinearities are small
corrections to Maxwell electrodynamics, they exhibit the same properties.
Three examples of Born-Infeld-like NLED are the Born-Infeld itself, the
Logarithmic and the Exponential electrodynamics.

\subsubsection{ Born-Infeld electrodynamics}

The Born-Infeld NLED was first proposed in 1934 by M. Born and L. Infeld
\cite{BornInfeld,BornInfeld1} and its Lagrangian is given by
\begin{equation}
\mathcal{L}_{BI}=b^{2}\left[  1-\sqrt{1-\frac{2F}{b^{2}}-\frac{G^{2}}{b^{4}}%
}\right]  . \label{BI Lagrangian}%
\end{equation}
This electrodynamic was created with the main purpose of avoiding the
divergence of a point-like particle self-energy, but it shows other
interesting features such as the absence of birefringence in vacuum
\cite{Boilat,CaLeoPJ}.

The electric displacement vector associated with (\ref{BI Lagrangian}) is
given by%
\begin{equation}
\vec{D}=\frac{\vec{E}+\frac{1}{b^{2}}\left(  \vec{E}\cdot\vec{B}\right)
\vec{B}}{\sqrt{1-\frac{E^{2}-B^{2}}{b^{2}}-\frac{\left(  \vec{E}\cdot\vec
{B}\right)  ^{2}}{b^{4}}}}. \label{DesBI}%
\end{equation}
In the weak field regime, $\mathcal{L}_{BI}$ can be approximated as%
\begin{equation}
\mathcal{L}_{BI}\approx F+\frac{1}{2b^{2}}\left(  F^{2}+G^{2}\right)  ,
\label{App LBI}%
\end{equation}
and (\ref{DesBI}) results in%
\[
D_{i}\approx\sum_{k}\varepsilon_{ki}E_{k},
\]
where%
\[
\varepsilon_{ki}=\delta_{ki}+\frac{1}{2b^{2}}\left(  E^{2}-B^{2}\right)
\delta_{ki}+\frac{1}{b^{2}}B_{i}B_{k},
\]
is the relative permittivity tensor. For the pure electrostatic case,%
\begin{equation}
\vec{D}\approx\left(  1+\frac{E^{2}}{2b^{2}}\right)  \vec{E}.
\label{DesEleApp}%
\end{equation}
The $\chi=E^{2}/2b^{2}$ term is identified as the electric susceptibility
which is associated with the medium's polarization. Note that, because $\chi>0$,
the vacuum behaves as a medium which resists to the formation of an electric field.

The displacement vector generated by the nucleus of a hydrogen-like atom (a
point particle system) is given by%
\begin{equation}
\vec{D}=\frac{Ze}{r^{2}}\hat{r}, \label{punctual charge}%
\end{equation}
where $e$ is the electron charge and $Z$ is the atomic number. The
substitution of this expression into (\ref{DesBI}), with $\vec{B}=0$, results
in the electric field given by%
\begin{equation}
\vec{E}_{BI}\left(  x\right)  =\frac{Z^{3}e}{a_{0}^{2}}\frac{1}{\sqrt
{x^{4}+\varepsilon^{4}}}\hat{r}, \label{BI electric field}%
\end{equation}
where $a_{0}$ is the Bohr radius and $x=Z\frac{r}{a_{0}}$ and $\varepsilon
=\sqrt{\frac{Z^{3}e}{a_{0}^{2}b}}$ are dimensionless parameters. The parameter
$\varepsilon$ measures how much the electric field deviates from Maxwell's electrodynamics.

\subsubsection{Logarithmic and Exponential electrodynamics}

The Logarithmic and Exponential electrodynamics belong to a special class,
called Born-Infeld-like NLED, which was proposed in order to study topics such
as inflation \cite{Altshuler} and exact solutions of spherically symmetric
static black holes \cite{Soleng,Hendi}. These electrodynamics are
characterized by having a finite self-energy solution for a point-like charge
but, unlike Born-Infeld NLED, they predict a birefringence effect in the
presence of an electromagnetic background field.

The Lagrangians for Logarithmic and Exponential NLED are given by
\cite{Helayel1,Helayel2}%
\begin{eqnarray}
L_{Lg}  &  = & -b^{2}\ln\left(  1-\frac{X}{b^{2}}\right)  ,\label{Lagrangian Log}%
\\
L_{Ex}  &  = & b^{2}\left(  e^{X/b^{2}}-1\right)  , \label{Lagrangian Exp}%
\end{eqnarray}
where $X=F+\frac{G^{2}}{2b^{2}}$. In the weak field limit, both Lagrangians
can be approximated by (\ref{App LBI}) and the electric displacement vectors
for pure electrostatic case are given by (\ref{DesEleApp}).

Following the same steps used in Born-Infeld case, we can calculate the
electric fields generated by the nucleus of a hydrogen-like atom:%
\begin{eqnarray}
\vec{E}_{Lg}\left(  x\right)   &  = & \frac{Z^{3}e}{\varepsilon^{4}a_{0}^{2}%
}\left(  \sqrt{x^{4}+2\varepsilon^{4}}-x^{2}\right)  \hat{r}%
,\label{Log electric field}\\
\vec{E}_{Ex}\left(  x\right)   &  = & \frac{Z^{3}e}{\varepsilon^{2}a_{0}^{2}%
}\sqrt{W\left(  \frac{\varepsilon^{4}}{x^{4}}\right)  }\hat{r},
\label{Exp electric field}%
\end{eqnarray}
where $x$ and $\varepsilon$ are defined as in (\ref{BI electric field}). The
function $W\left(  z\right)  $ is the Lambert function\footnote{For $z\in
\mathbb{R}
$ and $z\geq0$, $W\left(  z\right)  \geq0$ and it is monotonically increasing.
Besides, $\lim_{z\rightarrow0}\frac{W\left(  z\right)  }{z}=1$ and
$\lim_{z\rightarrow\infty}W\left(  z\right)  =\infty$.} defined as the inverse
function of $z\left(  W\right)  =We^{W}$. When $\varepsilon\rightarrow0$, both
electric fields reduce to the Maxwell case. Besides, $\vec{E}_{Ex}$ diverges
at the origin but slower than Maxwell, and $\vec{E}_{Lg}$ is bounded from
above in a similar way such as Born-Infeld field.

The behavior of the electric fields (\ref{BI electric field}),
(\ref{Log electric field}) and (\ref{Exp electric field}) are shown in figure
\ref{fig1}:%

\begin{figure}[th]
\includegraphics[height=5.5cm, width=8.8cm]{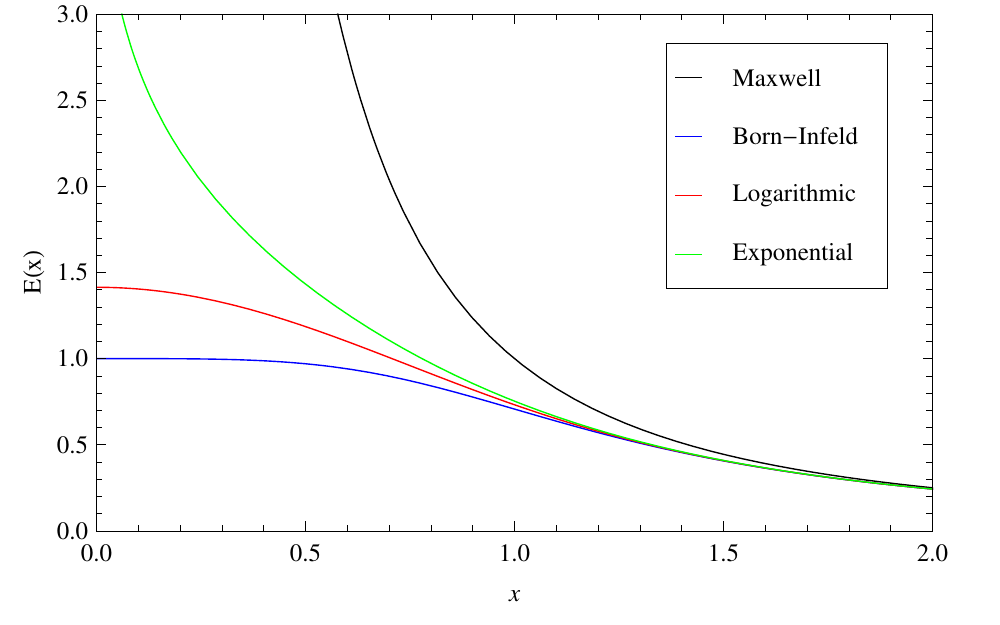}
\caption{Plot of $\left\vert \vec{E}\right\vert =E$ in units
$\frac{Z^{3}e}{a_{0}^{2}}$ adopting $\varepsilon=1$ versus distance $x$ for
Maxwell, Born-Infeld, Logarithmic and Exponential electrodynamics.}%
\label{fig1}%
\end{figure}

\section{Testing NLED using Hydrogen's Ionization Energy\label{sec - HIE}}

The theory about the energy levels of a hydrogen-like atom is described by the
quantization of Dirac equation and subject to several correction factors such
as the relativistic-recoil of the nucleus, electron self-energy, vacuum
polarization due to the creation of virtual electron-positron pairs, etc (for
details see \cite{RMP2002,PhyRep2001} and references therein). This
theoretical structure in the context of Maxwell electrodynamics establishes a
theoretical experimental agreement for the hydrogen's ionization (HI) energy
of $2$ parts per $10^{10}$ \cite{Kramida}. Thus, any correction to HI energy
generated by modifications in the Maxwell potential must be a small correction
and it can be treated perturbatively.

The Hamiltonian for a hydrogen-like atom in the context of NLED is given by%
\[
\hat{H}=\underset{\hat{H}_{0}}{\underbrace{\hat{K}+\hat{V}_{M}}}%
+\underset{\hat{H}_{p}}{\underbrace{\hat{V}_{G}-\hat{V}_{M}}},
\]
where $\hat{K}$ is the kinetic term, $\hat{V}_{M}$ is the Maxwell potential
and $\hat{V}_{G}$ is the potential energy of the NLED. Thus, $\hat{H}_{0}$ is
the usual hydrogen atom Hamiltonian and $\hat{H}_{p}$ is a perturbation of
this Hamiltonian.

The first order correction for HI energy\ due a Hamiltonian $\hat{H}_{p}$ is
given by%
\begin{equation}
E_{HI_{1}}=\left\langle \Psi_{100}\right\vert \hat{H}_{p}\left\vert \Psi
_{100}\right\rangle =4\left(  \frac{Z}{a_{0}}\right)  ^{3}%
{\displaystyle\int\limits_{0}^{\infty}}
drr^{2}e^{-\frac{2Zr}{a_{0}}}H_{p}\left(  r\right)  ,\label{EHI1}%
\end{equation}
where $\Psi_{100}$ is the ground state wave function%
\[
\Psi_{100}=\frac{1}{\sqrt{\pi}}\left(  \frac{Z}{a_{0}}\right)  ^{3/2}%
e^{-\frac{Zr}{a_{0}}},
\]
and%
\[
H_{p}\left(  r\right)  =V_{G}\left(  r\right)  -V_{M}\left(  r\right)  =-e%
{\displaystyle\int\limits_{r}^{\infty}}
\left[  E_{G}\left(  r^{\prime}\right)  -E_{M}\left(  r^{\prime}\right)
\right]  dr^{\prime}.
\]
$E_{G}\left(  r\right)  $ is the electric field absolute value generated by a
NLED and $E_{M}\left(  r\right)  =Ze/r^{2}$.

Defining the dimensionless variables $r=\frac{a_{0}y}{Z}$ and $r^{\prime
}=\frac{a_{0}}{Z}x$, expression (\ref{EHI1}) is rewritten as%
\begin{eqnarray}
E_{HI_{1}}  & = &\frac{4a_{0}e}{Z}%
{\displaystyle\int\limits_{0}^{\infty}}
dyy^{2}e^{-2y}%
{\displaystyle\int\limits_{y}^{\infty}}
\left[  E_{M}\left(  x\right)  -E_{G}\left(  x\right)  \right]  dx\label{ENI1v1}\\
& = &\frac{a_{0}e}{Z}%
{\displaystyle\int\limits_{0}^{\infty}}
dx\left(  E_{M}-E_{G}\right)  \left[  1-e^{-2x}\left(  1+2x+2x^{2}\right)
\right]  .\nonumber%
\end{eqnarray}
Since we are working in a perturbative regime where $E_{G}$ provides small
corrections to Maxwell's case we might be tempted to expand $E_{G}$ into a
Laurent series and to keep only the first correction term. This approximation,
however, is not valid at the lower limit of the integral since all terms
neglected become relevant as $x\leq1$. This is a crucial point in the
determination of $E_{HI_{1}}$ because this implies that each electrodynamic
will produce a different correction even though, in the weak field limit, they
are identical.

The term $E_{M}\left(  x\right)  $ in (\ref{ENI1v1}) can be explicitly worked
out and the $E_{HI_{1}}$ results in%
\begin{equation}
E_{HI_{1}}=\frac{Z^{2}e^{2}}{a_{0}}-\frac{a_{0}e}{Z}%
{\displaystyle\int\limits_{0}^{\infty}}
E_{G}\left[  1-e^{-2x}\left(  1+2x+2x^{2}\right)  \right]  dx.\label{CorrEn}%
\end{equation}
This expression will be the starting point to calculate the correction to the
hydrogen's ionization energy.

\subsection{Hydrogen's ionization energy for Born-Infeld electrodynamics}

The substitution of Born-Infeld electric field (\ref{BI electric field}) in
(\ref{CorrEn}) leads to%
\begin{equation}
E_{HI_{1}}^{BI}=\frac{Z^{2}e^{2}}{a_{0}}\left(  1-I_{1}^{BI}+I_{2}^{BI}%
+I_{3}^{BI}+I_{4}^{BI}\right)  ,\label{ECBI}%
\end{equation}
with%
\begin{eqnarray*}
I_{1}^{BI}  & = &%
{\displaystyle\int\limits_{0}^{\infty}}
\frac{dx}{\sqrt{x^{4}+\varepsilon^{4}}}=\frac{4\Gamma\left(  \frac{5}%
{4}\right)  ^{2}}{\sqrt{\pi}\varepsilon},\\
I_{2}^{BI}  & = &%
{\displaystyle\int\limits_{0}^{\infty}}
\frac{e^{-2x}dx}{\sqrt{x^{4}+\varepsilon^{4}}}=\frac{\varepsilon}{16\sqrt
{2}\pi^{2}}G_{15}^{51}\left(  \left.  \frac{\varepsilon^{4}}{16}\right\vert
_{-\frac{1}{2},-\frac{1}{4},-\frac{1}{4},0,\frac{1}{4}}^{\frac{1}{4}}\right)
,\\
I_{3}^{BI}  & = &%
{\displaystyle\int\limits_{0}^{\infty}}
\frac{2xe^{-2x}dx}{\sqrt{x^{4}+\varepsilon^{4}}}=\frac{\varepsilon^{2}}%
{8\sqrt{2}\pi^{2}}G_{15}^{51}\left(  \left.  \frac{\varepsilon^{4}}%
{16}\right\vert _{-\frac{1}{2},-\frac{1}{2},-\frac{1}{4},0,\frac{1}{4}}%
^{0}\right)  ,\\
I_{4}^{BI}  & = &%
{\displaystyle\int\limits_{0}^{\infty}}
\frac{2x^{2}e^{-2x}dx}{\sqrt{x^{4}+\varepsilon^{4}}}=\frac{\varepsilon^{3}%
}{8\sqrt{2}\pi^{2}}G_{15}^{51}\left(  \left.  \frac{\varepsilon^{4}}%
{16}\right\vert _{-\frac{3}{4},-\frac{1}{2},-\frac{1}{4},0,\frac{1}{4}%
}^{-\frac{1}{4}}\right)  ,
\end{eqnarray*}
where $G_{pq}^{mn}\left(  \left.  z\right\vert _{\vec{b}_{q}}^{\vec{a}_{p}%
}\right)  $ are the MeijerG functions \cite{Gradshteyn}.

For small corrections to Maxwell's potential $\varepsilon\ll1$, the
MeijerG functions can be approximated by\footnote{The
$\gamma_{E}$ is the Euler-Mascheroni constant.}%
\begin{eqnarray*}
I_{2}^{BI} & \approx &-2+2\gamma_{E}+\ln\left(  2\varepsilon^{2}\right)
+\frac{8\pi^{3/2}}{\Gamma\left(  -\frac{1}{4}\right)  ^{2}\varepsilon}
- \frac{4\pi^{3/2}}{\Gamma\left(  \frac{1}{4}\right)  ^{2}}\varepsilon \\
&& +\frac{2}{3}\varepsilon^{2}-\frac{\pi^{3/2}}{16\Gamma\left(
\frac{7}{4}\right)  ^{2}}\varepsilon^{3},\\
I_{3}^{BI} & \approx & -2\gamma_{E}-\ln\left(  2\varepsilon^{2}\right)
+\frac{8\pi^{3/2}}{\Gamma\left(  \frac{1}{4}\right)  ^{2}%
}\varepsilon-2\varepsilon^{2}+\frac{\pi^{3/2}}{4\Gamma\left(  \frac{7}%
{4}\right)  ^{2}}\varepsilon^{3},\\
I_{4}^{BI} & \approx &1-\frac{\pi^{3/2}}{\Gamma\left(  \frac{1}%
{4}\right)  \Gamma\left(  \frac{5}{4}\right)  }\varepsilon+2\varepsilon^{2}+\frac
{2\pi^{3/2}}{\Gamma\left(  -\frac{1}{4}\right)  \Gamma\left(
\frac{7}{4}\right)  }\varepsilon^{3}.
\end{eqnarray*}
Thus, the two first corrections for HI energy due the Born-Infeld
electrodynamics are given by
\begin{equation}
E_{HI_{1}}^{BI}\approx\left[  \frac{2}{3}\varepsilon^{2}-\frac{1}{3}\frac
{\pi^{\frac{3}{2}}}{\Gamma\left(  \frac{3}{4}\right)  ^{2}}\varepsilon
^{3}\right]  \frac{\left(  Ze\right)  ^{2}}{a_{0}},\label{EABI}%
\end{equation}
with $\varepsilon=\sqrt{\frac{Z^{3}e}{a_{0}^{2}%
b}}$. It is noteworthy that the first term in expression (\ref{EABI}) was
first obtained in \cite{Heller}. A positive $E_{HI_{1}}^{BI}$ indicates a
reduction in the ionization energy. This is consistent with a susceptibility
$\chi>0$ which reduces the value of the electric field generated by the nucleus.

Comparison of the numerical results for $\frac{a_{0}}{\left(  Ze\right)  ^{2}%
}E_{HI_{1}}^{BI}$ and the leading order approximation for different values of
$\varepsilon$ is shown in table $1$.
\begin{equation*}%
\centering
\begin{tabular}
[c]{|c|c|c|c|}\hline
$\varepsilon$ & numerical & first correction & relative error\\\hline
$10^{-1}$ & $5.66\times10^{-3}$ & $6.67\times10^{-3}$ & $17.758$ $\%$\\\hline
$10^{-3}$ & $6.65\times10^{-7}$ & $6.67\times10^{-7}$ & $0.185$ $\%$\\\hline
$10^{-5}$ & $6.67\times10^{-11}$ & $6.67\times10^{-11}$ & $0.0019$
$\%$\\\hline
\end{tabular}
\ \ \ \
\end{equation*}
TABLE 1: Results of $\frac{a_{0}}{\left(  Ze\right)  ^{2}}E_{HI_{1}}%
^{BI}$ for $\varepsilon=10^{-1}$, $10^{-3}$ and $10^{-5}$. The second column
shows the numerical result calculated from (\ref{ECBI}), the third column
shows the first order correction given by the first term in (\ref{EABI}) and
the last column shows the relative error between the two approaches.

\subsection{Hydrogen's ionization energy for Logarithmic electrodynamics}

The correction for the HI energy due the Logarithmic NLED is obtained using
the electric field (\ref{Log electric field}) in (\ref{CorrEn}):%
\begin{equation}
E_{HI_{1}}^{Lg}=\frac{Z^{2}e^{2}}{a_{0}\varepsilon^{4}}\left(  \varepsilon
^{4}-I_{1}^{Lg}-I_{2}^{Lg}-I_{3}^{Lg}-I_{4}^{Lg}\right)  ,\label{ECL}%
\end{equation}
with%
\begin{eqnarray*}
I_{1}^{Lg}  & = &%
{\displaystyle\int\limits_{0}^{\infty}}
\left(  \sqrt{x^{4}+2\varepsilon^{4}}-x^{2}\right)  dx=\frac{3\varepsilon
^{3}\Gamma\left(  -\frac{3}{4}\right)  ^{2}}{16\sqrt[4]{2}\sqrt{\pi}},\\
I_{2}^{Lg}  & = &%
{\displaystyle\int\limits_{0}^{\infty}}
\left(  \sqrt{x^{4}+2\varepsilon^{4}}-x^{2}\right)  e^{-2x}dx\\
& = &\frac{1}{4}+\frac{2^{\frac{3}{4}}\varepsilon}{4\pi^{2}}G_{15}^{51}\left(
\left.  \frac{\varepsilon^{4}}{8}\right\vert _{-\frac{1}{4},\frac{1}{2}%
,\frac{3}{4},1,\frac{5}{4}}^{\frac{5}{4}}\right)  ,\\
I_{3}^{Lg}  & = &%
{\displaystyle\int\limits_{0}^{\infty}}
\left(  \sqrt{x^{4}+2\varepsilon^{4}}-x^{2}\right)  e^{-2x}2xdx\\
& = &\frac{3}{4}+\frac{\varepsilon^{2}}{\pi^{2}}G_{15}^{51}\left(  \left.
\frac{\varepsilon^{4}}{8}\right\vert _{-\frac{1}{2},\frac{1}{2},\frac{3}%
{4},1,\frac{5}{4}}^{1}\right)  ,\\
I_{4}^{Lg}  & = &%
{\displaystyle\int\limits_{0}^{\infty}}
\left(  \sqrt{x^{4}+2\varepsilon^{4}}-x^{2}\right)  e^{-2x}2x^{2}dx\\
& = &\frac{3}{2}+\frac{2^{\frac{1}{4}}\varepsilon^{3}}{\pi^{2}}G_{15}%
^{51}\left(  \left.  \frac{\varepsilon^{4}}{8}\right\vert _{-\frac{3}{4}%
,\frac{1}{2},\frac{3}{4},1,\frac{5}{4}}^{\frac{3}{4}}\right)  .
\end{eqnarray*}

In the limit $\varepsilon\ll1$, the expressions above can be approximated by%
\begin{eqnarray*}
I_{2}^{Lg} & \approx &\frac{2^{\frac{3}{4}}\pi^{\frac{3}{2}}%
}{\Gamma\left(  -\frac{1}{4}\right)  \Gamma\left(  \frac{7}{4}\right)
}\varepsilon^{3}+\left(  \frac{5}{2}-2\gamma_{E}-\frac{1}{2}\ln\left(  8\varepsilon
^{4}\right)  \right)  \varepsilon^{4}\\
&& + \frac{\sqrt[4]{2}\pi^{\frac{3}{2}}}{\Gamma\left(  \frac
{5}{4}\right)  \Gamma\left(  \frac{9}{4}\right)  }\varepsilon^{5}-\frac{4\sqrt{2}%
}{9}\varepsilon^{6},\\
I_{3}^{Lg} & \approx &\left(  2\gamma_{E}-\frac{1}{2}+\frac{1}{2}\ln\left(
8\varepsilon^{4}\right)  \right)  \varepsilon^{4}
- \frac{\pi^{\frac{3}{2}}}{2^{\frac{3}{4}}\Gamma\left(
\frac{5}{4}\right)  \Gamma\left(  \frac{9}{4}\right)  }\varepsilon^{5} \\
&& +\frac{4\sqrt{2}}{3}\varepsilon^{6},\\
I_{4}^{Lg} & \approx &-\varepsilon^{4}+\frac{2^{\frac{1}{4}}\pi^{\frac{3}{2}%
}}{\Gamma\left(  \frac{1}{4}\right)  \Gamma\left(  \frac{9}%
{4}\right)  }\varepsilon^{5}-\frac{4\sqrt{2}}{3}\varepsilon^{6}.
\end{eqnarray*}
Thus, up to leading order, expression (\ref{ECL}) results in%
\begin{equation}
E_{HI_{1}}^{Lg}\approx\frac{\left(  Ze\right)  ^{2}}{a_{0}}\frac{4\sqrt{2}}%
{9}\varepsilon^{2}=\frac{4\sqrt{2}}{9}\frac{Z^{5}e^{3}}{a_{0}^{3}%
b}.\label{EAL}%
\end{equation}
This result is very similar to the first order Born-Infeld correction
(\ref{EABI}) differing only by a numerical factor of $\mathcal{O}\left(
1\right)  $.

Comparison of the numerical results for $\frac{a_{0}}{\left(  Ze\right)  ^{2}%
}E_{HI_{1}}^{Lg}$ and the leading order approximation is presented in table
$2$.%
\begin{equation*}
\begin{tabular}
[c]{|c|c|c|c|}\hline
$\varepsilon$ & numerical & first correction & relative error\\\hline
$10^{-1}$ & $5.325\times10^{-3}$ & $6.285\times10^{-3}$ & $18.044$
$\%$\\\hline
$10^{-3}$ & $6.274\times10^{-7}$ & $6.285\times10^{-7}$ & $0.188$ $\%$\\\hline
$10^{-5}$ & $6.285\times10^{-11}$ & $6.285\times10^{-11}$ & $0.0019$
$\%$\\\hline
\end{tabular}
\ \ \
\end{equation*}
TABLE 2: Results of $\frac{a_{0}}{\left(  Ze\right)  ^{2}}E_{HI_{1}}%
^{Lg}$ for $\varepsilon=10^{-1}$, $10^{-3}$ and $10^{-5}$. The second column
shows the numerical result calculated from (\ref{ECL}), the third column shows
the first order correction given by (\ref{EAL}) and the last column shows the
relative error between the two approaches.

\subsection{Hydrogen's ionization energy for Exponential electrodynamics}

The substitution of the Exponential NLED electric field
(\ref{Exp electric field}) in (\ref{CorrEn}) leads to%
\begin{equation}
E_{HI_{1}}^{Ex}=\frac{Z^{2}e^{2}}{a_{0}\varepsilon^{2}}\left(  \varepsilon
^{2}-I_{1}^{Ex}+I_{2}^{Ex}+2I_{3}^{Ex}+2I_{4}^{Ex}\right)  ,\label{ECE}%
\end{equation}
where%
\begin{equation}
I_{1}^{Ex}=%
{\displaystyle\int\limits_{0}^{\infty}}
\sqrt{W\left(  \frac{\varepsilon^{4}}{x^{4}}\right)  }dx=\frac{\sqrt{2}}%
{2}\Gamma\left(  \frac{1}{4}\right)  \varepsilon,\label{I1Exp}%
\end{equation}
and%
\begin{eqnarray*}
I_{2}^{Ex}  & = &%
{\displaystyle\int\limits_{0}^{\infty}}
\sqrt{W\left(  \frac{\varepsilon^{4}}{x^{4}}\right)  }e^{-2x}dx\text{, \ }\\
I_{3}^{Ex}  & = &%
{\displaystyle\int\limits_{0}^{\infty}}
\sqrt{W\left(  \frac{\varepsilon^{4}}{x^{4}}\right)  }e^{-2x}2xdx\text{, \ }\\
I_{4}^{Ex}  & = &%
{\displaystyle\int\limits_{0}^{\infty}}
\sqrt{W\left(  \frac{\varepsilon^{4}}{x^{4}}\right)  }e^{-2x}2x^{2}dx\text{.}%
\end{eqnarray*}
Integral $I_{1}^{Ex}$ was calculated using the properties of the Lambert
function $W$ after the change of variable $ue^{u}=\frac{\varepsilon^{4}}%
{x^{4}}$. The other three integrals do not have analytical solutions. However,
approximated solutions can be achieved following the steps described in
appendix \ref{sec - AppSol}. The leading order correction for HI energy due
the Exponential electrodynamics is obtained substituting (\ref{I1Exp}),
(\ref{I2Exp}), (\ref{I3Exp}) and (\ref{I4Exp}) into (\ref{ECE}):%
\begin{equation}
E_{HI_{1}}^{Ex}\approx\frac{\left(  Ze\right)  ^{2}}{a_{0}}\frac{\sqrt{\pi}%
}{3}\varepsilon^{2}=\frac{\sqrt{\pi}}{3}\frac{Z^{5}e^{3}}{a_{0}^{3}%
b}.\label{EAE}%
\end{equation}

Comparison of the numerical results for $\frac{a_{0}}{\left(  Ze\right)  ^{2}%
}E_{HI_{1}}^{Ex}$ and the leading order approximation is presented in table
$3$.%
\begin{equation*}
\begin{tabular}
[c]{|c|c|c|c|}\hline
$\varepsilon$ & numerical & first correction & relative error\\\hline
$10^{-1}$ & $4.989\times10^{-3}$ & $5.908\times10^{-3}$ & $18.424$
$\%$\\\hline
$10^{-3}$ & $5.897\times10^{-7}$ & $5.908\times10^{-7}$ & $0.193$ $\%$\\\hline
$10^{-5}$ & $5.908\times10^{-11}$ & $5.908\times10^{-11}$ & $0.0019$
$\%$\\\hline
\end{tabular}
\ \ \ \ \ \
\end{equation*}
TABLE 3: Results of $\frac{a_{0}}{\left(  Ze\right)  ^{2}}E_{HI_{1}}%
^{Ex}$ for $\varepsilon=10^{-1}$, $10^{-3}$ and $10^{-5}$. The second column
shows the numerical result calculated from (\ref{ECE}), the third column shows
the first order correction given by (\ref{EAE}) and the last column shows the
relative error between the two approaches.

\subsection{Constraining parameter $b$}

The ground state energy level correction calculated in the previous sections
is generically given by%
\[
E_{HI_{1}}=K\frac{Z^{5}e^{3}}{3a_{0}^{3}b},
\]
where $K=2$, $4\sqrt{2}/3$ and $\sqrt{\pi}$ for the Born-Infeld, Logarithmic
and Exponential electrodynamics respectively. The experimental value of
hydrogen atom ionization energy in frequency units is \cite{Kramida}%
\[
\nu=3288086856.8\pm0.7 \text{\, MHz.}%
\]
It is important to emphasize that this value measured by the National
Institute of Standards and Technology (NIST) is a purely experimental result
which does not assume any theoretical background. The same does not occur with
other measurements available in the literature - e.g. Particle Data Group
\cite{PDG} - which provide the ionization energy already assuming Maxwell's
electrostatic potential.

Imposing that the energy correction must be smaller than $3$ times the
experimental error $\sigma_{\nu}$ i.e. $E_{HI_{1}}<3h\sigma_{\nu}$, parameter
$b$ (with $Z=1$) is constrained by the expression%
\begin{equation}
b>K\frac{e^{3}}{9a_{0}^{3}h\sigma_{\nu}}. \label{b}%
\end{equation}
Restoring SI units and using values given by \cite{CODATA2014} we obtain%
\begin{equation}
b>5.37\times10^{20}K\frac{V}{m},
\label{b value}
\end{equation}
which in terms of the dimensionless parameter $\varepsilon$ corresponds to
\begin{equation}
\varepsilon<\frac{3.\,\allowbreak1\times10^{-5}}{\sqrt{K}}. \label{eps value}%
\end{equation}
The last result is consistent with the approximation $\varepsilon\ll1$ used in
the previous theoretical calculations.

For the particular Born-Infeld case, the expression (\ref{b value}) results in%
\begin{equation}
b_{BI}>1.07\times10^{21}\frac{V}{m}. \label{b BI}%
\end{equation}

Historically, the first estimation for $b_{BI}$ was done by M. Born and L.
Infeld in 1934 \cite{BornInfeld} relating in an oversimplified manner the mass
of the electron with its self-energy. The value found by those authors was
$b_{BI}>1.2\times10^{20}\frac{V}{m}$. Forty years later G. Soft et al.
\cite{Soff} obtained $b_{BI}>1.7\times10^{22}\frac{V}{m}$ through a
theoretical-experimental comparison involving muonic spectral transitions in
lead atoms $_{82}Pb$. Although an order of magnitude more precise than
(\ref{b BI}), the theoretical modeling presented in \cite{Soff} is
questionable because it does not take into account the loss of spherical
symmetry due to the presence of the remaining leptons. This kind of approach
is particularly problematic in NLED where the loss of spherical symmetry
implies in $\vec{\nabla}\times\vec{D}\neq0$ \cite{CarleyKiess} and
consequently invalidates the expression (\ref{D and E}) used in \cite{Soff}.
More recently in the 21st century it was suggested by J. M. D\'{a}vila et al.
\cite{Davila2014} that $b_{BI}$ can be bound from the magnetars spectrum due
to the effect of photon splitting. Following this approach, the authors of
\cite{Davila2014} estimated $b_{BI}>2.0\times10^{19}\frac{V}{m}$. Finally, at
the end of 2016 ATLAS collaboration announced the first direct measurement of
photon-photon scattering in ultra-peripheral heavy-ion collisions \cite{Atlas
2016,Atlas 2017}. Based on this measure, J. Ellis et al. \cite{Ellis}
constrained Born-Infeld parameter\ to $b_{BI}>4.3\times10^{27}\frac{V}{m}$.
This last result is six orders of magnitude larger than (\ref{b BI}), but it
was obtained from a much more complex theoretical-experimental arrangement
\cite{KlusekGawenda2016} and therefore subject to greater uncertainty. In this
sense, the treatment adopted here provides a simpler laboratory, and a
mathematical method adaptable without difficulty to a great variety of NLED
such as, for instance, the Logarithmic and Exponential electrodynamics.

\section{Final Remarks\label{sec - Final}}

In this work the ground state energy level correction $E_{HI_{1}}$ for the
hydrogen atom generated by three Born-Infeld-like electrodynamics was
obtained. More specifically, a general expression for the correction
$E_{HI_{1}}$ was derived through a perturbative approach, then this correction
was calculated for the Born-Infeld, Logarithmic and Exponential
electrodynamics. Using the experimental uncertainty for HI energy, the free
parameters $b$'s of each of these NLED were lower bounded, and for the
particular Born-Infeld case the result found was compared with other
constraints present in the literature. It is worth mentioning that the method
developed here based on the expression (\ref{CorrEn}) and the techniques of
the appendix \ref{sec - AppSol} can easily be extended to constrain other
nonlinear electrodynamics.

An important point in the derivation of $E_{HI_{1}}$ concerns the need to know
the electric field exactly (see discussion below Eq. (\ref{ENI1v1})). This
point can be observed by the distinct values obtained for $E_{HI_{1}}^{BI}$,
$E_{HI_{1}}^{Lg}$ and $E_{HI_{1}}^{Ex}$. Although different, these values are
similar and we can wonder if the expression (\ref{b value}) could be used to
constrain a more general class of NLED. The necessary and sufficient condition
to apply the result (\ref{b value}) to others NLED is related to the behavior
of the electric field. Observing figure \ref{fig1} and the values of $K$ ($K_{BI}=2$,
$K_{Lg}=4\sqrt{2}/3$ and $K_{Ex}=\sqrt{\pi}$) we see that the greater is the
difference between the Maxwell and Born-Infeld-like NLED electric fields the
higher is the $K$ value. Thus, we can state that any NLED whose electric field
absolute value $E_{NLED}$ fulfills the condition $E_{BI}<E_{NLED}<E_{Ex}$ will
have $b_{Ex}<b_{NLED}<b_{BI}$. Also, since $K$ slightly varies from $K_{Ex}$
to $K_{BI}$ we can estimate that any NLED which has an $E_{NLED}$ near to
$E_{BI}$ or $E_{Ex}$ will have its free parameter bounded by $b_{NLED}%
\gtrsim10^{21}V/m$. Thus, we can impose limits on a broad class of NLED only
by knowing the behavior of its electric field.

Finally, it is important to discuss the possibility of application involving
the electrodynamics of Euler-Heisenberg (EH) \cite{HeisenbergEuler,Dunne2004}.
EH electrodynamics is an effective description of the self-interaction process
due the electron-positron virtual pairs present in QED (vacuum polarization).
Thus, starting from EH NLED one could think of using the procedure developed
in this work to obtain, in an alternative way, the vacuum polarization
correction for the hydrogen's ionization energy \cite{Uehling,RMP2002}. The
problem with this approach is that the EH electrodynamics is built assuming a
slowly varying electromagnetic field in distances of the order of the electron
Compton wavelength $\lambda_{e}$, and this requirement is not satisfied in the
calculation of $E_{HI_{1}}$. The essential part of the integral $E_{HI_{1}}$
is in the range $[0$,$1[$ (see appendix \ref{sec - AppSol}), and within this
range the EH electric field rapidly varies at distances of order $\lambda_{e}%
$. Therefore, the vacuum polarization effect associated with the hydrogen atom
can not be described by the Euler-Heisenberg effective electrodynamics
\cite{Uehling}.

\begin{acknowledgements}
The authors acknowledge A.E. Kramida and R.R. Cuzinatto for their useful
comments. They are also grateful to CNPq-Brazil for financial support.
\end{acknowledgements}

\appendix

\section{$I_{2}^{Ex}$, $I_{3}^{Ex}$ and $I_{4}^{Ex}$ Approximate
Solutions\label{sec - AppSol}}

The first step to calculate $I_{2}^{Ex}$ is split the integral in the ranges
$[0,1)$ and $\left[  1,\infty\right)  $.%
\begin{equation}
I_{2}^{Ex}=\underset{A}{\underbrace{\int_{0}^{1}dx\sqrt{W\left(
\frac{\varepsilon^{4}}{x^{4}}\right)  }e^{-2x}}}+%
{\displaystyle\int\limits_{1}^{\infty}}
dx\sqrt{W\left(  \frac{\varepsilon^{4}}{x^{4}}\right)  }e^{-2x}.\label{Ap1}%
\end{equation}
As $\varepsilon\ll1$ (small corrections to Maxwell's case), the $W$ function
can be approximated by%
\[
W\left(  \frac{\varepsilon^{4}}{x^{4}}\right)  \approx\frac{\varepsilon^{4}%
}{x^{4}}+\mathcal{O}\left(  \frac{\varepsilon^{8}}{x^{8}}\right),
\]
which for the second integral is a great approximation and thus results in%
\[
{\displaystyle\int\limits_{1}^{\infty}}
dx\sqrt{W\left(  \frac{\varepsilon^{4}}{x^{4}}\right)  }e^{-2x}=\varepsilon
^{2}\left(  \frac{1}{e^{2}}+2\operatorname{Ei}_{1}\left(  -2\right)  \right),
\]
where
\[
\operatorname{Ei}_{n}\left(  x\right)  \equiv\int_{1}^{\infty}%
e^{-xt}/t^{n}dt,
\]
is the exponential integral function.

The next step is to work out with the first integral. Performing the variable
substitution $ue^{u}=\frac{\varepsilon^{4}}{x^{4}}$, the integral $A$ leads
to:%
\begin{eqnarray*}
A  &  = &-\frac{\varepsilon}{4}%
{\displaystyle\int\limits_{\infty}^{W\left(  \varepsilon^{4}\right)  }}
\left(  u^{-\frac{3}{4}}+u^{\frac{1}{4}}\right)  e^{-\frac{u}{4}%
}e^{-2\varepsilon u^{-\frac{1}{4}}e^{-\frac{u}{4}}}\\
&  \approx &\frac{\varepsilon}{4}%
{\displaystyle\int\limits_{\varepsilon^{4}}^{\infty}}
\left(  u^{-\frac{3}{4}}+u^{\frac{1}{4}}\right)  e^{-\frac{u}{4}}\sum
_{n=0}^{\infty}\frac{1}{n!}\left(  -2\varepsilon u^{-\frac{1}{4}}e^{-\frac
{u}{4}}\right)  ^{n},
\end{eqnarray*}
where the exponential of exponential was expanded in a Taylor series. The term
$e^{-\frac{u}{4}}$ ensures the convergence of the integral at the limit
$u\rightarrow\infty$ when $n=0$. It is important to emphasize that although
$\varepsilon\ll1$ the sum can not be truncated in the first terms. This occurs
because the lower bound of integration depends on $\varepsilon$. Thus, all
terms of the sum will contribute to $\varepsilon^{2}$, $\varepsilon^{3}$, etc.

The third step is to rewrite $A$ in terms of exponential integral functions
and expanding these functions up to order $\varepsilon^{4}$:%

\begin{eqnarray*}
A  & \approx &\frac{\varepsilon}{4}\sum_{n=0}^{\infty}\frac{\left(  -2\right)
^{n}}{n!}\varepsilon^{n}%
{\displaystyle\int\limits_{\varepsilon^{4}}^{\infty}}
\left(  u^{-\frac{n+3}{4}}+u^{-\frac{n-1}{4}}\right)  e^{-\frac{n+1}{4}u}\\
& \approx &\sum_{n=0}^{\infty}\frac{\left(  -2\right)  ^{n}}{4\left(  n!\right)
}\left[  \varepsilon^{6}\operatorname{Ei}_{\frac{n-1}{4}}\left(  \frac{n+1}%
{4}\varepsilon^{4}\right)  \right. \\
&& +\left.  \varepsilon^{2}\operatorname{Ei}_{\frac{n+3}{4}}\left(  \frac
{n+1}{4}\varepsilon^{4}\right)  \right] \\
& \approx & \sum_{n=0}^{3}\frac{\left(  -2\right)  ^{n}}{n!}\left[
2^{\frac{-\left(  1+n\right)  }{2}}\left(  1+n\right)  ^{\frac{n-5}{4}}%
\Gamma\left(  \frac{1-n}{4}\right)  \varepsilon^{n+1}\right]  \\
&& +\sum_{n=0}^{\infty}\frac{\left(  -2\right)  ^{n}}{n!\left(  n-1\right)
}\varepsilon^{2}+\mathcal{O}\left(  \varepsilon^{5}\right)
\end{eqnarray*}

The first two terms in the r.h.s. of $A$ above cancel out for $n=1$ (although
they separately diverge). This can be seen by expanding $\Gamma\left(
\frac{1-n}{4}\right)  $ around $n=1$,%
\[
\Gamma\left(  \frac{1-n}{4}\right)  =-\frac{4}{n-1}-\gamma_{E}+\mathcal{O}%
\left(  n-1\right)  .
\]

Thus,%
\begin{eqnarray*}
A  & \approx &\underset{n=0}{\underbrace{\frac{1}{\sqrt{2}}\Gamma\left(
\frac{1}{4}\right)  \varepsilon-\varepsilon^{2}}}+\underset{n=1}%
{\underbrace{\left(  \frac{1}{2}\gamma_{E}-1+\frac{1}{2}\ln\frac
{\varepsilon^{4}}{2}\right)  \varepsilon^{2}}}\\
&& +\varepsilon^{2}\left(  -\frac{1}{e^{2}}-1+2\gamma_{E}-2\operatorname{Ei}%
_{1}\left(  -2\right)  +\ln4\right)  \\
&& +\sum_{n=2}^{3}\frac{\left(  -2\right)  ^{n}}{n!}\left(  2^{\frac{-1-n}{2}%
}\left(  1+n\right)  ^{\frac{n-5}{4}}\Gamma\left(  \frac{1-n}{4}\right)
\varepsilon^{n+1}\right)  ,
\end{eqnarray*}
where it was used the relation%
\[
\sum_{n=2}^{\infty}\frac{\left(  -2\right)  ^{n}}{n!\left(  n-1\right)
}=-\frac{1}{e^{2}}-1+2\gamma_{E}-2\operatorname{Ei}_{1}\left(  -2\right)
+\ln4.
\]

By substituting $A$ into (\ref{Ap1}) we obtain the final expression for
$I_{2}^{Ex}$:%

\begin{eqnarray}
I_{2}^{Ex}  & \approx &\frac{\sqrt{2}}{2}\Gamma\left(  \frac{1}{4}\right)
\varepsilon+\left(  \frac{5}{2}\gamma_{E}-3+\frac{1}{2}\ln8\varepsilon
^{4}\right)  \varepsilon^{2}\nonumber\\
&& + \frac{1}{3^{\frac{3}{4}}\sqrt{2}}\Gamma\left(  -\frac{1}{4}\right)
\varepsilon^{3}+\frac{\sqrt{\pi}}{3}\varepsilon^{4}.\label{I2Exp}%
\end{eqnarray}

The computation procedure for the integrals $I_{3}^{Ex}$ and $I_{4}^{Ex}$
follows the same steps described above. For $I_{3}^{Ex}$ we have:
\begin{eqnarray*}
I_{3}^{Ex}  & = &%
{\displaystyle\int\limits_{0}^{\infty}}
dx\sqrt{W\left(  \frac{\varepsilon^{4}}{x^{4}}\right)  }e^{-2x}x\\
& \approx &\underset{B}{\underbrace{%
{\displaystyle\int\limits_{0}^{1}}
dx\sqrt{W\left(  \frac{\varepsilon^{4}}{x^{4}}\right)  }e^{-2x}x}}%
-\varepsilon^{2}\operatorname{Ei}_{1}\left(  -2\right)  .
\end{eqnarray*}
Using $\frac{\varepsilon^{4}}{x^{4}}=ue^{u}$, the integral $B$ is rewritten as%
\begin{eqnarray*}
B  & \approx &\frac{\varepsilon^{2}}{4}\sum_{n=0}^{\infty}\frac{\left(
-2\right)  ^{2}}{n!}\varepsilon^{n}%
{\displaystyle\int\limits_{\varepsilon^{4}}^{\infty}}
\left(  u^{-1}+1\right)  u^{-\frac{n}{4}}e^{-\frac{\left(  n+2\right)  }{4}%
u}\\
& \approx &\sum_{n=0}^{\infty}\frac{\left(  -2\right)  ^{n}}{4\left(  n!\right)
}\left[  \varepsilon^{2}\operatorname{Ei}_{\frac{n}{4}+1}\left(  \frac{n+2}%
{4}\varepsilon^{4}\right)  \right.  \\
&& +\left.  \varepsilon^{6}\operatorname{Ei}_{\frac{n}{4}}\left(  \frac{n+2}%
{4}\varepsilon^{4}\right)  \right]  \\
& \approx &\sum_{n=0}^{2}\frac{\left(  -2\right)  ^{n}}{n!}\left[
2^{-1-\frac{n}{2}}\left(  2+n\right)  ^{\frac{n}{4}-1}\Gamma\left(  -\frac
{n}{4}\right)  \varepsilon^{2+n}\right]  \\
&& +\sum_{n=0}^{\infty}\frac{\left(  -2\right)  ^{n}}{\left(  n!\right)
n}\varepsilon^{2}+\mathcal{O}\left(  \varepsilon^{5}\right)  \\
& \approx &\left(  \frac{1}{2}-\frac{5}{4}\gamma_{E}-\frac{1}{4}\ln
8\varepsilon^{4}+\operatorname{Ei}\left(  -2\right)  \right)  \varepsilon
^{2}\\
& -&\frac{1}{3^{\frac{3}{4}}\sqrt{2}}\Gamma\left(  -\frac{1}{4}\right)
\varepsilon^{3}-\frac{\sqrt{\pi}}{2}\varepsilon^{4}.
\end{eqnarray*}
Thus,%
\begin{eqnarray}
I_{3}^{Ex}  & \approx &\left(  \frac{1}{2}-\frac{5}{4}\gamma_{E}-\frac{1}{4}%
\ln8\varepsilon^{4}\right)  \varepsilon^{2}\nonumber\\
&& - \frac{1}{3^{\frac{3}{4}}\sqrt{2}}\Gamma\left(  -\frac{1}{4}\right)
\varepsilon^{3}-\frac{\sqrt{\pi}}{2}\varepsilon^{4}.\label{I3Exp}%
\end{eqnarray}
And for the integral $I_{4}^{Ex}$ we have:%
\begin{eqnarray*}
I_{4}^{Ex}  & =& \int_{0}^{\infty}dx\sqrt{W\left(  \frac{\varepsilon^{4}}{x^{4}%
}\right)  }e^{-2x}x^{2}\\
& \approx &\underset{C}{\underbrace{%
{\displaystyle\int\limits_{0}^{1}}
dx\sqrt{W\left(  \frac{\varepsilon^{4}}{x^{4}}\right)  }e^{-2x}x^{2}}}%
+\frac{\varepsilon^{2}}{2e^{2}}.
\end{eqnarray*}
Once more using the substitution $\frac{\varepsilon^{4}}{x^{4}}=ue^{u}$, the
integral $C$ is rewritten as:%
\begin{eqnarray*}
C  & \approx &\frac{\varepsilon^{3}}{4}\sum_{n=0}^{\infty}\frac{\left(
-2\right)  ^{n}}{n!}\varepsilon^{n}\int_{\varepsilon^{4}}^{\infty}\left(
u^{-\frac{n+5}{4}}+u^{-\frac{n+1}{4}}\right)  e^{-\left(  \frac{n+3}%
{4}\right)  u}\\
& \approx &\sum_{n=0}^{\infty}\frac{\left(  -2\right)  ^{n}}{4n!}\left[
\varepsilon^{2}\operatorname{Ei}_{\frac{n+5}{4}}\left(  \frac{n+3}%
{4}\varepsilon^{4}\right)  \right.  \\
&& +\left.  \varepsilon^{6}\operatorname{Ei}_{\frac{n+1}{4}}\left(  \frac
{n+3}{4}\varepsilon^{4}\right)  \right]  \\
& \approx &\sum_{n=0}^{1}\frac{\left(  -2\right)  ^{n}}{n!}\left[
2^{\frac{-3-n}{2}}\left(  3+n\right)  ^{\frac{n-3}{4}}\Gamma\left(
-\frac{n+1}{4}\right)  \varepsilon^{3+n}\right]  \\
&& +\sum_{n=0}^{\infty}\frac{\left(  -2\right)  ^{n}}{\left(  n!\right)
\left(  n+1\right)  }\varepsilon^{2}+\mathcal{O}\left(  \varepsilon
^{5}\right)  \\
& \approx &\left(  -\frac{1}{2e^{2}}+\frac{1}{2}\right)  \varepsilon^{2}%
+\frac{1}{3^{\frac{3}{4}}\sqrt{8}}\Gamma\left(  -\frac{1}{4}\right)
\varepsilon^{3}+\frac{\sqrt{\pi}}{2}\varepsilon^{4}%
\end{eqnarray*}
Thus,%
\begin{equation}
I_{4}^{Ex}\simeq\frac{1}{2}\varepsilon^{2}+\frac{1}{3^{\frac{3}{4}}\sqrt{8}%
}\Gamma\left(  -\frac{1}{4}\right)  \varepsilon^{3}+\frac{\sqrt{\pi}}%
{2}\varepsilon^{4}.\label{I4Exp}%
\end{equation}

Results (\ref{I2Exp}), (\ref{I3Exp}) and (\ref{I4Exp}) are necessary to obtain
equation (\ref{EAE}) appearing in the main text.


\begin{thebibliography}{99}                                                                                               %


\bibitem {BornInfeld}M. Born, L. Infeld, Proc. R. Soc. Lond. A \textbf{144},
852 (1934).

\bibitem {BornInfeld1}M. Born, L. Infeld, Proc. R. Soc. Lond. A \textbf{147},
522 (1934).

\bibitem {FraTse}E.S. Fradkin, A.A. Tseytlin, Phys. Lett. B \textbf{163}, 123 (1985).

\bibitem {HeisenbergEuler}W. Heisenberg, H. Euler, Z. Phys. \textbf{98}, 714 (1936).

\bibitem {Dunne2004}G.V. Dunne, From Fields to Strings: Circumnavigating
Theoretical Physics, Ian Kogan Memorial Collection, 2004.

\bibitem {Altshuler}B.L. Altshuler, Class. Quantum Grav. \textbf{7}, 189 (1990).

\bibitem {Soleng}H. Soleng, Phys. Rev. D \textbf{52}, 6178 (1995).

\bibitem {Hendi}S.H. Hendi, J. High Energy Phys. \textbf{03}, 065 (2012).

\bibitem {Hendi2} S.H. Hendi et al.,  Eur. Phys. J. C \textbf{76}, 150 (2016).

\bibitem {Plebanski}J. Plebansky, Lectures on Non-linear Electrodynamics,
Nordita, Copenhagen, 1968.

\bibitem {BH1}E. Ayon-Beato, A. Garcia, Phys. Rev. Lett. \textbf{80}, 5056 (1998).

\bibitem {BH4}J. Diaz-Alonso, D. Rubiera-Garcia, Phys. Rev. D \textbf{81},
064021 (2010).

\bibitem {BH5}J. Diaz-Alonso, D. Rubiera-Garcia, Phys. Rev. D \textbf{82},
085024 (2010).

\bibitem {BH6}R. Ruffini, Y.-B. Wu, S.-S. Xue, Phys. Rev. D \textbf{88},
085004 (2013).

\bibitem {BH7}R.R. Cuzinato, C.A.M. de Melo, K.C. de Vasconcelos, L.G.
Medeiros, P.J. Pompeia, Astrophys. Space Sci \textbf{359}, 59 (2015).

\bibitem {BH8}  S.H. Hendi, B. Eslam Panah, S. Panahiyan, J. High Energy Phys. \textbf{11}, 157 (2015). 

\bibitem {BH9} S.H. Hendi et al.,  Eur. Phys. J. C \textbf{76}, 571 (2016).

\bibitem {Cosm1}V.A. De Lorenci, R. Klippert, M. Novello, J.M. Salim, Phys.
Rev. D \textbf{65}, 063501 (2002).

\bibitem {Cosm2}V.V. Dyadichev, D.V. Galtsov, A.G. Zorin, M.Y. Zotov, Phys.
Rev. D \textbf{65}, 084007 (2002).

\bibitem {Cosm4}M. Novello, S.E.P. Bergliaffa, J. Salim, Phys. Rev. D
\textbf{69}, 127301 (2004).

\bibitem {Cosm5}M. Novello, A.N. Araujo, J.M. Salim, Int. J. of Mod. Phys. A
\textbf{24}, 5639 (2009).

\bibitem {Cosm6}L.G. Medeiros, Int. J. of Mod. Phys. D \textbf{21}, 1250073 (2012).

\bibitem {CaLeoPJ}C.A.M. de Melo, L.G. Medeiros, P.J. Pompeia, Mod. Phys.
Lett. A \textbf{30}, 1550025 (2015).

\bibitem {Boilat}G. Boillat, J. Math. Phys. \textbf{11}, 941 (1970).

\bibitem{Fouche}M. Fouch\'{e}, R. Battesti, C. Rizzo, Phys. Rev. D
\textbf{93}, 093020 (2016). 

\bibitem {Soff}G. Soff, Phys. Rev. A \textbf{7}, 903 (1973).

\bibitem {CarleyKiess}H. Carley, M.K.-H. Kiessling, Phys. Rev. Lett.
\textbf{96}, 030402 (2006).

\bibitem {Atlas 2017}ATLAS Collaboration, Nat. Phys. \textbf{13}, 852 (2017).

\bibitem {Ellis}J. Ellis, N. Mavromatos, T. You, Phys. Rev. Lett.
\textbf{118}, 261802 (2017).

\bibitem {Davila2014}J.M. Davila, C. Schubert, M.A. Trejo, Int. J. Mod. Phys.
A \textbf{29}, 1450174 (2014).

\bibitem {PVLAS}F. Della Valle et al., Eur. Phys. J. C. \textbf{76}, 24 (2016).\

\bibitem {Helayel1}P. Gaete, J. Helay\"{e}l-Neto, Eur. Phys. J. C.
\textbf{74}, 2816 (2014).

\bibitem {Helayel2}P. Gaete, J. Helay\"{e}l-Neto, Eur. Phys. J. C.
\textbf{74}, 3182 (2014).

\bibitem {RMP2002}P.J. Mohr, B.N. Taylor, Rev. Mod. Phys. \textbf{77}, 1 (2005).

\bibitem {PhyRep2001}M.I. Eides, H. Grotch, V.A. Shelyuto, Phys. Rep.
\textbf{342}, 63 (2001).

\bibitem {Kramida}A.E. Kramida, A critical compilation of experimental data on
spectral lines and energy levels of hydrogen, deuterium, and tritium, Atomic
Data and Nuclear Data Tables, 96(6), 586-644, 2010.

\bibitem {Gradshteyn}D. Zwillinger, Table of Integrals, Series, and Products,
Elsevier Science, 2014.

\bibitem {Heller}G. Heller, L. Motz, Phys. Rev. \textbf{46}, 502 (1934).

\bibitem {PDG} K.A. Olive et al. (Particle Data Group), Chin. Phys. C
\textbf{38}, 090001 (2014).

\bibitem {CODATA2014}P.J. Mohr, D.B. Newell, B.N. Taylor, Rev. Mod. Phys.
\textbf{88}, 035009 (2016).

\bibitem {Atlas 2016}ATLAS Collaboration, ATLAS-CONF-2016-111, 2016.

\bibitem {KlusekGawenda2016}M. K\l usek-Gawenda, P. Lebiedowicz, A. Szczurek,
Phys. Rev. C \textbf{93}, 044097 (2016).

\bibitem {Uehling}E.A. Uehling, Phys. Rev. \textbf{48}, 55 (1935).
\end{thebibliography}
\end{document}